\documentclass[12pt]{iopart}
 \usepackage{graphicx}

\usepackage{palatino}
 \usepackage[all]{xy}
    \SelectTips{eu}{10 }     %use the nicer arrowheads
    \everyxy={<2.5em,0em>:} % Sets the scale I like
 \usepackage{fancyhdr}      %%%%%%%%%%%% Pagestyle stuff %%%%%%%%%%%%%%%
\usepackage{srcltx}
\usepackage{eufrak,mathrsfs,bbm,pxfonts,parskip}
\newcommand\beq{\begin{equation}}
\newcommand\eeq{\end{equation}}
\newcommand\bea{\begin{eqnarray}}
\newcommand\eea{\end{eqnarray}}
\newcommand{\SR}{\left( \pm \right)}
\newcommand{\SP}{\left( + \right)}
\newcommand{\SM}{\left( - \right)}
\DeclareMathAlphabet{\mathpzc}{OT1}{pzc}{m}{it}
\begin{document}
\title{2+1 Quantum Gravity with Barbero-Immirzi like parameter on Toric Spatial Foliation} 
\author{Rudranil Basu and Samir K Paul}
\address{S. N. Bose National Centre 
for Basic Sciences, JD Block, Sector III, Salt Lake City, \phantom{iiiiii} Kolkata 
700098, India. }
\ead{rudranil@bose.res.in , smr@bose.res.in}
\begin{abstract}
We consider  gravity in 2+1 space-time  dimensions, with negative cosmological constant and a `Barbero-Immirzi' (B-I) like  parameter, when the space-time topology is  of  the  form  $ T^2 \times \mathbbm{R}$. The  phase  space  structure, both  in covariant  and  canonical  framework  is  analyzed.  Full  quantization  of  the theory  in  the  'constrain first'  approach  reveals  a  finite  dimensional  physical  Hilbert  space. An  explicit  construction  of  wave  functions  is presented. The  dimension  of  the  Hilbert  space  is  found  to  depend  on  the `Barbero-Immirzi'  like  parameter  in  an  interesting  fashion. Comparative study of this parameter in light of some of the recent findings in literarure for similar theories is presented. 
\end{abstract}
\section{Introduction}
Quantum gravity in 2+1 dimensions have been an object of serious research for quite some time. A very special feature of this theory is that the theory of 2+1 gravity in first order formalism (with/without) cosmological constant can be written as a gauge theory \cite{Witten:1988hc}. Although there is no propagating mode in this theory \cite{Deser:1984ab}, it is striking that it admits of a CFT at the boundary when the theory is considered in an asymptotically AdS space-time \cite{Brown:1986qj}. On the other hand, the idea of incorporating local degrees of freedom (gravitons) is quite old \cite{Deser:1982vy} and recenly there has been proliferations through newer avenues (new massive gravity) without \cite{Bergshoeff:2009hq} or (cosmological new massive gravity) with cosmological constant \cite{Liu:2009bk}, which we will hereafter refer collectively as topologically massive gravity (TMG). Gravity in 2+1 space-time even without graviton modes took an intresting turn after existence of black hole solutions was ensured \cite{Banados:1992wn}. Subsequent important works in the context of AdS/CFT correspondence \cite{Strominger:1997eq}, \cite{Birmingham:1998jt} warrants the importance of this model.

Even if one restricts oneself with 2+1 gravity models without propagating degrees of freedom, quantization of the theory poses a non-trivial problem in its own right in the sense that one has to study this problem by quantizing the phase space keeping in mind that topology of the spacetime would play an important role in deciding the quantum theory. If the phase space is finite dimensional one can do away with problems regarding renormalizability even in the non-perturbative regime \cite{Witten:1988hc},\cite{Witten:2007kt}. In this work we would deal with 2+1 gravity (with negative cosmological constant) described by vielbeins (triads) and $SO(2,1)$ connections on a (pseudo)Riemann-Cartan manifold which is not asymptotically AdS~and aim to compare results with asymptotically AdS calculations already available in literature. The bulk theory of 2+1 gravity which will not be taken as TMG in our case, can be expressed as Chern Simons theory with non-compact gauge groups, choice of the group being strictly determined by the cosmological constant \cite{Witten:1988hc}. Whereas 2+1 Chern-Simons theory with compact gauge group gives natural explanations for many constructions in Conformal Field Theory (CFT)\cite{Verlinde:1989hv}, \cite{Witten:1988hf}, the same theory with non-compact gauge group (in the interest of gravity) can give rise to significant generalizations of these constructions \cite{Witten:1989ip}. That there is only global degree of freedom, can be seen both in the metric formulation and in the triad-connection formulation of general relativity. All the local degrees of freedom are frozen when the constraints, all of which in this case (namely the Gauss, the Hamiltonian and the Diffeomorphism) are first class, are imposed. The global degrees of freedom turn out to be finite in number if there is no topologically non-trivial boundary in the space time\cite{Nelson:1989zd}.

In the case of negative cosmological constant the Chern-Simons action corresponding to 2+1 gravity (hereafter referred as CSG) can be written as  an $SO(2,1) \times SO(2,1)$ gauge theory \cite{Witten:1988hc}, which is purely topological as opposed to TMGs. The topology of the physical phase space of the theory being nontrivial  one  has  to  take  recourse  to geometric quantization  \cite{Woodhouse:1980pa}. This formulation of quantization in the 'constrain first' approach  was studied for an SL(2,R) Chern-Simons theory with rational charges \cite{Imbimbo:1991he} where a finite dimensional Hilbert space was constructed on the almost torus part of the physical phase space and it was argued that the Hilbert space on the total phase space would be finite dimensional. Spatial slice in this case was chosen to be a torus. General quantization procedure of Chern Simons theories in the realm of geometric quantization was exhaustively studied in \cite{Axelrod:1989xt}.

More generalized versions of CSG retaining its topological nature came into prominence through the works of Mielke {\it et al} \cite{Mielke:1991nn}, \cite{Baekler:1992ab}. In the present work we consider a special case of such generalized CSG \cite{Cacciatori:2005wz} with a negative cosmological constant and a new parameter which immitates the Barbero Immirzi parameter of 3+1 gravity\cite{Holst:1995pc}; the possibility of this generalization also was hinted in the pioneering work \cite{Witten:1988hc}. The gauge group for the corresponding CSG still remains $SO(2,1) \times SO(2,1)$. We discuss the quantization in the 'constrain first' approach where one first solves the classical constraints and then attempts to quantize the resulting phase space of gauge invariant variables. It was revealed in \cite{Elitzur:1989nr} that this approach fails to incorporate the `shift' in the central charge of the current algebra of the Wess-Zumino-Witten conformal field theory. Nevertheless, as explained in \cite{Imbimbo:1991he} that the above difficulty is overcome as one of the inequivalent Hilbert spaces has exactly the unitary structure of the vector space of the current blocks of the Wess-Zumino-Witten theory. 

We also wish to point out that in a later work \cite{Ezawa:1994nv}, an explicit parameterization of the physical phase space for CSG on toric spatial foliation and negative cosmological constant was done. There, in contrast to geometric quantization, the phase space was modified to a suitable cotangent bundle by a surgery of the non-trivial phase space and trivializing its topology. Conventional procedure of canonical quantization was carried out in that modified phase space. No comment however on the dimensionality of the Hilbert space was made.

In section \ref{Phase_space} we construct the phase space for CSG with the negative cosmological constant and the Barbero-Immirzi like parameter $\gamma$. We note that the original CSG action with negative cosmological constant can be written as a sum of two $SO(2,1)$ Chern Simons actions with equal and opposite signed levels. The difference in present case follows by adding a different Lagrangian (with arbitrary coefficient) to the original one, which also gives same equation of motion. Due to this modification the Chern-Simons levels (for each of the two $SO(2,1)$ sectors) carrying the footprints of both the negative cosmological constant and the new B-I like parameter can be tuned. In the analysis (of similar TMGs) pertaining to asymptotically AdS calculations \cite{Li:2008dq}, \cite{Grumiller:2010rm} this freedom is crucial, in the sense that one of the Chern Simons levels can be tuned to zero, which corresponds to the `chiral limit' of the dual CFT. We, in our present work also try to make sense of this limit. The physical phase space is the moduli space of flat gauge connections, modulo gauge transformations on our choice of spatial foliation, which is genus-1 Riemann surface. It turns out to be a torus punctured at a point (may be chosen to be origin) with a plane also punctured at a point (also chosen to be origin of the plane) and glued to the torus  through a closed curve ($S^1$) around the origin (common puncture) the plane being $Z_2$ folded through the origin .

In section \ref{geo_q}  we discuss the geometric quantization of the phase space \cite{Woodhouse:1980pa}. A complete basis for the physical Hilbert space is  constructed in terms of theta functions. Note that due to introduction of this new parameter only both the Chern-Simons levels can be adjusted to be positive and rational. During quantization this becomes important since the dimensionality of the Hilbert space of the quantized theory is directly related with these levels.  The corresponding charge in the CSG is no longer an integer owing to the fact that the Weil's integrality condition on the Chern-Simons charge disappears as a consequence of the non compactness of the gauge group  which in our case is $SO(2,1)$ \cite{Witten:1988hf},\cite{Elitzur:1989nr}. A discussion on the restrictions on physical parameters coming from the quantization is also presented and compared with those from \cite{Witten:2007kt}.

\section{Phase space of 2+1 gravity with Immirzi like parameter}\label{Phase_space}
In this section we will demonstrate the classical covariant phase space (actually a pre-symplectic manifold) \cite{cov_phs_spc} of 2+1 gravity with a negative cosmological constant and an Immirzi like parameter. The canonical phase space (non-covariant), the Hamiltonian structure, the constraint analysis and gauges relevant to the theory will also be discussed.
\subsection{Holst-like 2+1 Gravity as a Chern Simons theory}
Action for 2+1 gravity with negative cosmological constant $\Lambda = -\frac{1}{l^2}$ on a space time manifold $M$ in first order formalism is 
\bea \label{Palatini}
I_{GR}= \frac{1}{8 \pi G} \int _{M} e^I \wedge \left(2 d \omega _I + \epsilon _I{}^{JK} \omega _J \wedge \omega _K + 
\frac{1}{3l^2}\epsilon _I{}^{JK} e_J \wedge e_K \right)
\eea
$e^I$ are the $SO(2,1)$ orthonormal triad frame and the $\omega ^I$ are connections (or canonically projected local connection) of the frame-bundle with structure group $SO(2,1)$.% which become spin connection once the torionless equation of motion is satisfied, ie on-shell. Since we will consider space of solutions in this paper, $\omega ^I$ will always qualify as spin connection.
     The above action is well defined and differentiable in absence of boundaries. Although in presence of boundary (internal and/or asymptotic)  \cite{ash} one has to add suitable boundary terms to the action in order to have a finite action with well defined (differentiable) variation.

The equations of motion for space times without boundary are
\bea \label{EOM1}
%\begin{subequations}
F_I := 2d\omega _I +  \epsilon _I{}^{JK} \omega _J \wedge \omega _K = -\frac{1}{l^2} \epsilon _I{}^{JK} e_J \wedge e_K \\
\label{EOM2}
T_I := d e _I + \epsilon _{IJK}e^J \wedge \omega ^K = 0
%\end{subequations}
\eea
Note that this theory describes gravitational interaction as long as the triad system $e^I _a$ are invertible. ($a,b ..$ Latin indices are abstract space time indices). The connection 1-forms $\omega ^I _a$ qualify as the spin-connections if \eref{EOM2} is satisfied and the space-time becomes (pseudo)Riemann manifold as opposed to initial strucure of a (psudo)Riemann-Cartan manifold.

A more general model for 2+1 gravity with negative cosmological constant was introduced by Mielke {\it et al} \cite{Mielke:1991nn},\cite{Baekler:1992ab} and later studied extensively in \cite{Cacciatori:2005wz},\cite{Blagojevic:2003wn},\cite{Blagojevic:2005hd}, which without matter fields read:
\bea \label{mielke}
I = a I_1 +b I_2 + \alpha _3 I_3 + \alpha _4 I_4
\eea
where
\begin{eqnarray*} 
I_1 =\int _{M} e^I \wedge \left(2 d \omega _I + \epsilon _I{}^{JK} \omega _J \wedge \omega _K \right) \\
I_2 = \int _{M} \epsilon ^{IJK} e_I \wedge e_J \wedge e_K \\
I_3 = \int _{M} \omega^I \wedge d \omega _I +  \frac{1}{3} \epsilon _{IJK} \omega^I \wedge \omega^J \wedge\omega^K \\
I_4 = \int _{M} e^I \wedge d e_I + \epsilon _{IJK} \omega^I \wedge e^J \wedge e^K
\end{eqnarray*}
However this model does not reproduce the equations of motion \eref{EOM1} and \eref{EOM2} for arbitrary values of the parameters $ a,b, \alpha _3, \alpha _4$. We choose, as a special case of the above model, those values of these parameters which gives the expected equations of motion as in \cite{Witten:1988hc},\cite{Witten:2007kt},\cite{Cacciatori:2005wz}, \cite{Manschot:2007zb}-\cite{Maloney:2007ud}:
\bea \label{para_choice}
a=\frac{1}{8 \pi G} ~~ b = \frac{1}{24 \pi G l^2} ~~ \alpha _3 = \frac{l}{8 \pi G \gamma} ~~ \alpha _4 = \frac{1}{8 \pi G \gamma l}
\eea
$\gamma $ is introduced as new dimensionless parameter from 2+1 gravity perspective. Effectively \eref{para_choice} is the equation of a 3 dimensional hypersurface parametrized by $G,l, \gamma$ in the 4-d parameter space of $a,b, \alpha _3 , \alpha _4$.

It calls for a little digression for the Chern Simons formulation. Following \cite{Witten:1988hc},\cite{Cacciatori:2005wz} one introduces the $SO(2,1)$ or equivalently $SL(2, \mathbbm{R})$ or 
$SU(1,1)$ connections for a principal bundle over the same base space of the frame bundle:
\begin{eqnarray*}
A^{\SR I} := \omega ^I \pm \frac{e^I}{l} ~.
\end{eqnarray*}
It is easily verifiable that the action 
\bea \label{split}
\tilde I=\frac{l}{16\pi G} \left( I^{\SP} - I^{\SM}\right)
\eea
is same as \eref{Palatini} in absence of boundaries. 
Where
\bea \label{defn_I}
I^{\SR} = \int _{M}\left( A^{\SR I} \wedge d A^{\SR}_I + \frac{1}{3} \epsilon _{IJK} A^{\SR I} \wedge A^{\SR J} \wedge 
A^{\SR K} \right)
\eea
are two Chern Simons actions with gauge group $SO(2,1)$, the lie algebras being given by
\begin{eqnarray}
\left[J^{\SP}_I , J^{\SP}_J\right] = \epsilon _{IJK} J^{\SP K} && \quad  \left[J^{\SM}_I , J^{\SM}_J\right] = \epsilon _{IJK} J^{\SM K} \nonumber \\
&& \left[J^{\SP}_I , J^{\SM}_J\right] = 0.
\end{eqnarray}
 The metric on the Lie algebra is chosen to be 
$$\langle J^{\SR I}, J^{\SR J} \rangle = \frac{1}{2} \eta ^{IJ}$$
where $J^{\SR I}$ span the $SO(2,1)$ (or $SL(2, \mathbbm{R})$ or $SU(1,1)$) Lie algebras for the two theories.

One striking feature of this formulation is that the last two terms of \eref{mielke} can also be incorporated in terms of $ A ^{\SR}$, for ($ \alpha _3 = l^2 \alpha _4$) as:
$$I^{\SP} + I^{\SM} =2 \int _{M}\left(\omega^I \wedge d \omega _I + \frac{1}{l^2} e^I \wedge d e_I + \frac{1}{3} \epsilon _{IJK} \omega^I \wedge \omega^J \wedge\omega^K + \frac{1}{l^2}\epsilon _{IJK} \omega^I \wedge e^J \wedge e^K \right)$$
and the same equations of motion \eref{EOM1} and \eref{EOM2} are also found from varying this action. 

We thus propose the action 
\bea 
\label{new}
I &=&\frac{l}{16\pi G} \left( I^{\SP} - I^{\SM}\right) + \frac{l}{16\pi G \gamma} \left( I^{\SP} + I^{\SM}\right) 
\nonumber\\
&=&\frac{l}{16\pi G}\left[ \left(1/\gamma +1 \right) I^{\SP} + \left(1/\gamma -1 \right) I^{\SM}\right]
\eea
with a dimensionless non-zero coupling $\gamma$. This action \eref{new} upon variations with respect to $A^{\SP}$ and $A^{\SM}$ give equations of motion as expected from Chern Simons theories. This imply that the connections $A^{\SR}$ are flat:
\bea \label{flat}
\mathcal{F}^{\SR}_I : = dA^{\SR}_I + \epsilon_{IJK} A^{\SR J} \wedge A^{\SR K} = 0.
\eea 
It is also easy to check that the above flatness conditions of these $SO(2,1)$ bundles \eref{flat} are equivalent to the equations of motion of general relativity \eref{EOM1}, \eref{EOM2}. 

This is a good point to stop and probe into the physical relevance of this new parameter comparing with 3+1 dimensional gravity. In order to proceed we notice that the new action is in spirit very much like the Holst action \cite{Holst:1995pc} used in 3+1 gravity. In our case the parameter $\gamma$ can superficially be thought of being the 2+1 dimensional counterpart of the original Barbero-Immirzi parameter. Moreover the part $I^{\SP} + I^{\SM}$ of the action in this light qualifies to be at par with the topological (non-dynamical) term one adds with the usual Hilbert-Palatini action in 3+1 dimensions, since this term we added (being equal to a Chern Simons action for space-times we consider) is also non-dynamical. But more importantly the contrast is in the fact that the original action, which is dynamical in the 3+1 case is also non-dynamical here, when one considers local degrees of freedom only. 

Another striking contrast between the original B-I parameter and the present one lies in the fact that in the 3+1 scenario $\gamma$ parameterizes canonical transformations in the phase space of general relativity. From the canonical pair of the $SU(2)$ triad (time gauge fixed and on a spatial slice) and spin-connection one goes on finding an infinitely large set of pairs parameterized by $\gamma$. The connection is actually affected by this canonical transformation, and this whole set of parameterized connections is popularly known as the Barbero-Immirzi connection. The fact that this parameter induces canonical transformation can be checked by seeing that the symplectic structure remains invariant under the transformation on-shell. On the other hand for the case at hand, ie 2+1 gravity, as we will see in the following sub-section that inclusion of finite $\gamma$ is not a canonical transformation and it does not keep the symplectic structure invariant.

\subsection{Symplectic Structure on the Covariant Phase Space}
Consider a globally hyperbolic space-time manifold endowed neither with an internal nor an asymptotic boundary and let it allows foliations \footnote{On-shell \eref{EOM2} implies the space-time can be given (pseudo) Riemannian structure. With respect to the associated metric (0,2)-tensor and a time like vector field $t^a$ the manifold is assumed to be Cauchy-foliated.} $M \equiv \Sigma \times \mathbbm R $, with $\Sigma$ being compact and $ \partial\Sigma =0$. 

In view of \cite{cov_phs_spc} the covariant phase space, ie the space of solutions of the equations of motion the theory is $\mathcal V_{F}^{\SP}\times \mathcal V_{F}^{\SM}$, product of spaces of flat $SO(2,1)$ connections as discussed in the last section. We now intend to find the pre-symplectic structure \footnote{It is being called the pre-symplectic structure since as we will point out later that only on the constraint surfaces this has the property to be gauge invariant. When we have a phase space parameterized by gauge invariant variables, this pre-symplectic structure will induce a symplectic structure on that.}. For that purpose, we start with the Lagrangian 3-form that gives the above action:
\bea \label{lag}
L&=&\frac{l}{16\pi G} \left(1/\gamma + 1\right)\left( A^{\SP I} \wedge d A^{\SP}_I + \frac{1}{3} \epsilon _{IJK} A^{\SP I} \wedge A^{\SP J} \wedge 
A^{\SP K}\right)\nonumber \\ &+&\frac{l}{16\pi G}\left(1/\gamma - 1\right)\left( A^{\SM I} \wedge d A^{\SM}_I + \frac{1}{3} \epsilon _{IJK} A^{\SM I} \wedge A^{\SM J} \wedge A^{\SM K}\right) 
\eea

The standard variation gives on-shell:
$$\delta L =: d \Theta (\delta) = d \Theta ^{\SP}(\delta)+ d \Theta ^{\SM}(\delta)$$
where $\left(16\pi G/l\right)\Theta ^{\SR} (\delta) = \left(1/\gamma \pm 1\right)\delta A^{\SR I} \wedge A^{\SR}_I $.
%And the gauge invariant phase space is $\mathcal V_{F}^{\SP}\times \mathcal V_{F}^{\SM} / gauge ~ transformations $
The procedure of second variations \cite{cov_phs_spc} then gives the pre-symplectic current 
\bea
J(\delta _1 , \delta _2) &=&  J^{\SP}(\delta _1 , \delta _2)+J^{\SM}(\delta _1 , \delta _2) \nonumber\\ 
\mbox{ where }&&  J^{\SR}(\delta _1 , \delta _2) = 2\delta _{[1}\Theta ^{\SR}(\delta _{2]}) \nonumber                         
\eea
which is a closed 2-form ($d J(\delta _1 , \delta _2) = 0$) on-shell. The closure of $J$ and the fact that we are considering space-time manifolds which allow closed Cauchy foliations imply that the integral $\int _{\Sigma} J(\delta _1 , \delta _2)$ is actually foliation independent, ie independent of choice of $\Sigma$. Hence the expression $\int _{\Sigma} J(\delta _1 , \delta _2)$ is manifestly covariant and qualifies as the pre-symplectic structure on $\mathcal V_{F}^{\SP}\times \mathcal V_{F}^{\SM}$. We thus define: 
\bea \label{symp1}
\Omega = \Omega ^{\SP}+\Omega^{\SM}
\eea
where
\bea \label{symp2}
\Omega ^{\SR}\left( \delta _1, \delta _2\right) &=& \int _{\Sigma}J^{\SR}(\delta _1 , \delta _2)\nonumber\\
&=&\frac{l}{8\pi G} \left(1/\gamma \pm 1 \right)\int _{\Sigma} \delta _{[1} A^{\SR I} \wedge \delta _{2]} A^{\SR}_I
\eea
At this point we would like to note two important features of this symplectic structure:
\begin{itemize}
\item In the 3+1 case the extra contribution of the Holst term (with coeffecient $1/ \gamma$)in the symplectic structure can be shown to vanish on-shell. Hence it is guaranteed that in the covariant phase space $\gamma$ has the role of inducing canonical transformations. On the other hand, in the present case, it is very much clear from the above expression, that the $\gamma$ dependent term cannot vanish, as suggested in the previous subsection. So, what we have at hand are infinite inequivalent theories for 2+1 gravity each having different canonical structure and parameterized by different values of $\gamma$ at the classical level itself.

\item The other point worth noticing is that $ \Omega$ is indeed gauge invariant and it can be checked by choosing one of the two $ \delta$ s to produce infinitesimal $SO(2,1)$ gauge transformations or infinitesimal diffeomorphisms and keeping the other arbitrary. In both these cases $ \Omega \left( \delta_{SO(2,1)}, \delta \right)$ and $ \Omega \left( \delta_{\scriptsize{\mbox {diffeo}} } , \delta \right)$ vanish on the constraint surface, recognizing these two classes of vectors in the covariant phase space as the `gauge' directions.
\end{itemize}
\subsection{Canonical Phase Space}
From the above covariant symplectic structure one can instantly read off the following canonical equal-time (functions designating the foliations as level surfaces) Poisson brackets:
\bea \label{PB1}
\{A^{\SR I}_i (x,t),A^{\SR J}_j (y,t)\}= \frac{8\pi G /l}{1/\gamma \pm 1} \varepsilon _{ij} \eta ^{IJ} \delta ^2 \left(x,y \right)
\eea
where $\varepsilon _{ij}$ is the usual alternating symbol on $\Sigma$.

Interestingly in terms of the Palatini variables, the above Poisson bracket reads as
\bea \label{PB2}
\{\omega^{ I}_i (x,t),e^{ J}_j (y,t)\}&=& 4\pi G \frac{\gamma ^2}{\gamma ^2 -1} \varepsilon _{ij} \eta ^{IJ} 
\delta ^2 \left(x,y \right)\nonumber\\
\{\omega^{ I}_i (x,t),\omega^{ J}_j (y,t)\} &=& -4\pi G \frac{\gamma /l}{\gamma ^2 -1} \varepsilon _{ij} \eta ^{IJ} 
\delta ^2 \left(x,y \right) \\
\{e^{ I}_i (x,t),e^{ J}_j (y,t)\} &=& -4\pi G \frac{\gamma l}{\gamma ^2 -1} \varepsilon _{ij} \eta ^{IJ} \delta ^2 
\left(x,y \right) \nonumber
\eea
As expected in the limit $\gamma \rightarrow \infty$ the Poisson brackets reduce to those of usual Palatini 
theory:
\bea \label{PB3}
\{\omega^{ I}_i (x,t),e^{ J}_j (y,t)\}&=& 4\pi G  \varepsilon _{ij} \eta ^{IJ} \delta ^2 \left(x,y \right) \nonumber\\
\{\omega^{ I}_i (x,t),\omega^{ J}_j (y,t)\} &=& 0  \\
\{e^{ I}_i (x,t),e^{ J}_j (y,t)\} &=& 0 \nonumber
\eea

We here wish to concentrate on the Hamiltonian and the constraint structure of the theory. In terms of the Chern Simons gauge fields these are the $SO(2,1)$ Gauss constraints as illustrated below.
The Legendre transformation is done by space-time splitting of the action $I$ given by\eref{new}
\bea \label{Legendre}
\hspace{-2.3cm}
\frac{16\pi G }{l}I = (1/\gamma +1 ) \int _{\mathbbm R} dt \int _{\Sigma}d^2 x \varepsilon^{ij}\left(-A_i^{\SP I} \partial _0 A^{\SP}_{jI} + 2 A_0^{\SP I} \partial _i A^{\SP}_{jI} + \epsilon ^{IJK}A_{0I}^{\SP}A_{iJ}^{\SP}A_{jK}^{\SP}\right)\nonumber \\
\hspace{-1cm}+ (1/\gamma -1 ) \int _{\mathbbm R} dt \int _{\Sigma}d^2 x \varepsilon^{ij}\left(-A_i^{\SM I} \partial _0 A^{\SM}_{jI} + 2 A_0^{\SM I} \partial _i A^{\SM}_{jI} + \epsilon ^{IJK}A_{0I}^{\SM}A_{iJ}^{\SM}A_{jK}^{\SM}\right)
\eea
First terms in the integrands are kinetic terms and from them one can again extract \eref{PB1}. The Hamiltonian is given by 
$$\mathcal H = \mathcal H^{\SP}+\mathcal H^{\SM}$$
where
$$\mathcal H^{\SR} = \frac{l \left(1/\gamma \pm 1 \right)}{16\pi G}\varepsilon^{ij}\left(2 A_0^{\SR I} \partial _i A^{\SR}_{jI} + \epsilon ^{IJK}A_{0I}^{\SR}A_{iJ}^{\SR}A_{jK}^{\SR}\right)$$
The fields $A^{\SR}_{0I}$ are the Lagrange multipliers and we immediately have the primary constraints
\bea \label{constraint-1}
\mathcal G ^{\SR}_I = \frac{l\left(1/\gamma \pm 1\right)}{16\pi G} \varepsilon^{ij}\left( \partial _i A^{\SR}_{jI} + \frac{1}{2}\epsilon _I{}^{JK}A_{iJ}^{\SR}A_{jK}^{\SR}\right) \approx 0
\eea
Since $\mathcal H ^{\SR}= A_0 ^{\SR I} \mathcal G ^{\SR}_I \approx 0$ the Hamiltonian is therefore weakly zero. Again the primary constraint being proportional to the Hamiltonian, there are no more secondary constraints in the theory {\it a la} Dirac.
Now consider the smeared constraint
$$\mathcal G ^{\SR}\left(\lambda\right) = \int _\Sigma d^2 x \lambda ^I \mathcal G ^{\SR}_I$$
for some $\lambda = \lambda ^I J_I \in \mathfrak{so}(2,1)$ in the internal space. It now follows that this smeared constraints close among themselves:
\begin{eqnarray}\label{constraint0}
\{ \mathcal G ^{\SR}\left(\lambda\right),\mathcal G ^{\SR}\left(\lambda ^{\prime}\right)\} = \mathcal G ^{\SR}\left(\left[\lambda,\lambda ^{\prime}\right]\right)
\end{eqnarray}
and the $SO(2,1)$ Lie algebra is exactly implemented on the canonical phase space. Hence clearly these are the `Gauss' constraints generating $SO(2,1)$ gauge transformations separately for the $\SP$ type and the $\SM$ type gauge fields. The closure of these constraints on the other hand means that these are first class and there are no second class constraints. A close look on \eref{constraint-1} reveals that this constraint is nothing but vanishing of the gauge field curvature \eref{flat} when pulled back to $\Sigma$. The temporal component $A_{0I}$ is non-dynamical, being just a Lagrange multiplier. Hence all the dynamics of the theory determined by \eref{flat} is constrained as \eref{constraint-1}. Hence there is no local physical degree of freedom in the theory, which is related to the justified recognition of Chern Simons theories as `topological'. We now wish to probe in to implications of this constraint structure in the gravity side, the $\gamma \rightarrow \infty$ case of which was discussed by various authors, e.g. \cite{Witten:1988hc}.

The Legendre transformation now is carried out through the space-time split action \eref{Legendre} in terms of the variables pertaining relevance to the gravity counterpart of the theory as
\bea \label{legendre1}
\frac{16\pi G }{l}I &=&-2\int _{\mathbbm R} dt\int _{\Sigma} d^2 x \varepsilon^{ij}\underbrace{\left[1/\gamma\left(\omega _i ^I \partial _0 \omega _{jI}+\frac{1}{l^2}e _i ^I \partial _0 e _{jI}\right)+\frac{2}{l}\omega _i ^I \partial _0 e _{jI}\right]}_{\mbox{kinetic terms}}\nonumber \\ &{}&
+4\int _{\mathbbm R} dt\int _{\Sigma} d^2 x \varepsilon^{ij}\Bigg[\frac{1}{l}\left(\omega _0 ^I + \frac{1}{\gamma l}e_0 ^I\right)\left(\partial_i e_{jI}+\epsilon_I{}^{JK}\omega _{iJ}e_{jK}\right)\nonumber \\ &{}&
+\left(\frac{1}{\gamma}\omega _0 ^I + \frac{1}{l}e_0 ^I\right)\left(\partial_i \omega_{jI}+\frac{1}{2}\epsilon_I{}^{JK}\left(\omega _{iJ}\omega_{jK}+\frac{1}{l^2}e _{iJ}e_{jK}\right)\right)\Bigg]
\eea
One then envisages the part save the kinetic part as the Hamiltonian in the units of $\frac{l}{16\pi G}$ and $\left(\omega _0 ^I + \frac{1}{\gamma l}e_0 ^I\right)$, $\left(\frac{1}{\gamma}\omega _0 ^I + \frac{1}{l}e_0 ^I\right)$ as the Lagrange multipliers with the following as the constraints, after suitable rescaling\footnote{A rescaling with the factor $ \frac{1- \gamma ^2}{ \gamma ^2}$ is done in order to avoid apparent divergences in the constraint algebra at the points $ \gamma \rightarrow \pm 1$}:
\bea \label{constraint2}
P^I := \frac{1- \gamma ^2}{8\pi G \gamma ^2} \varepsilon^{ij}\left(\partial_i e_{j}^I+\epsilon ^I{}_{JK}\omega _i^Je_j^K\right) \approx 0 \nonumber\\
S^I := \frac{l (1-\gamma ^2)}{8\pi G \gamma ^2}\varepsilon^{ij}\left(\partial_i \omega_j^I+\frac{1}{2}\epsilon ^I{}_{JK}\left(\omega _i^J\omega_j^K+\frac{1}{l^2}e _i^Je_j^K\right)\right) \approx 0
\eea
Let us define their smeared versions as
$$P( \lambda) := \int _{ \Sigma} d^2 x \lambda ^I P _I \mbox{ and } S( \lambda) := \int _{ \Sigma} d^2 x \lambda ^I S _I$$
for $ \lambda \in \mathfrak{so}(2,1)$.
One can also check the expected closure of the constraint algebra of $S$ and $P$ which guarantees their first class nature:
\begin{eqnarray} \label{constraint3}
\{S( \lambda), S ( \lambda ^{ \prime})\} &=& \gamma ^{-1} S([ \lambda, \lambda ^{\prime}])- P([ \lambda, \lambda ^{\prime}]) \nonumber\\
\{S( \lambda), P ( \lambda ^{ \prime})\} &=& - S([ \lambda, \lambda ^{\prime}])+ \gamma ^{-1} P([ \lambda, \lambda ^{\prime}]) \\
\{P( \lambda), P ( \lambda ^{ \prime})\} &=& \gamma^{-1} S([ \lambda, \lambda ^{\prime}])-  P([ \lambda, \lambda ^{\prime}])\nonumber
\end{eqnarray}
Linear combinations of $\omega _0^I$ and $e_0^I$ are Lagrange multipliers and hence these fields themselves are non dynamical. We thus infer that all the dynamical informations through the equations of motion \eref{EOM1}, \eref{EOM2} are encoded in the constraints \eref{constraint2}. In the limit $\gamma \rightarrow \infty$, as e.g. in \cite{Witten:1988hc} $P$ generate local Lorentz ie, $SO(2,1)$ Lorentz transformations and $S$ generate diffeomorphisms for the frame variables. Since in finite $ \gamma$ case too these are first class, one should expect them to generate some gauge transformation. To see changes brought in by the modified symplectic structure we first compute the transformations induced by these constraints:
\begin{eqnarray}\label{def_trans1}
\{e_i ^I (x,t), P(\lambda)\} = -\frac{l}{2} \left[\gamma ^{-1}  \underbrace{\left(\partial _ i \lambda ^I + \epsilon ^{IJK} \omega _{iJ} \lambda _K\right)}_{D_i \lambda ^I} + \frac{1}{l} \epsilon ^{IJK}\lambda _J e_{iK}\right] \\
\{\omega _i ^I (x,t), P(\lambda)\} = \frac{1}{2}\left[  D_i \lambda ^I + \frac{1}{l \gamma} \epsilon ^{IJ}{}_K \lambda _J e^K _i\right]
\end{eqnarray}
The infinitesimal local $SO(2,1)$ Lorentz transformations, ie $e \rightarrow e + \lambda \times e, \quad \omega \rightarrow \omega + d \lambda + \lambda \times \omega$ are seen to be successfully generated by $P( \lambda)$ in the limit $ \gamma \rightarrow \infty$. But for finite $ \gamma$, the transformations are deformed in a sense that infinitesimal diffeomorphisms are also generated along with Lorentz transformations. Similarly the Lie transports generated by the diffeomorphism generator $S$ are also deformed due to the modified symplectic structure as:
\begin{eqnarray}\label{def_trans2s}
\{e_i ^I (x,t), S(\lambda)\} = \frac{l}{2}\left[  D_i \lambda ^I + \frac{1}{l \gamma} \epsilon ^{IJ}{}_K e_{Ji} \lambda ^K\right] \\
\{\omega _i ^I (x,t), S(\lambda)\} = -\frac{1}{2}\left[ \gamma ^{-1} D_i \lambda ^I + \frac{1}{l} \epsilon ^{IJ}{}_K \lambda _J e^K_i \right] 
\end{eqnarray} 
In this case we also notice that the usual diffeomorphism generator is generating local Lorentz transformations for finite $ \gamma$. We sure can find suitable linear combinations of these two generators which separately and purely generate local Lorentz and diffeomorphisms.

The striking difference between roles of the original 3+1 Barbero Immirzi paramter and the present $ \gamma$ can again be envisaged in terms of the usual ADM canonical pairs: the spatial metric $ h_{ij}$ and the dual momentum $\pi ^{ij}= \sqrt{h} \left( K^{ij}-h^{ij}K\right)$, where $K^{ij}$ is the extrinsic curvature and $K$ is its trace. Using $h_{ij}= g_{ij} = e_i^I e_{jI}$, we have:
$$ \{h_{ij}(x,t),h_{kl}(y,t)\} = -4\pi G \frac{\gamma}{\gamma ^2 -1} \left( \varepsilon _{ik} h_{jl}+\varepsilon _{il} h_{jk}+\varepsilon _{jk} h_{il}+\varepsilon _{jl} h_{ik}\right)\delta ^2(x,y).$$
Similar Poisson brackets involving $\pi ^{ij}$ can also be calculated, which are more cumbersome. The point we get across from this bracket, is that while components of the spatial metric Poisson commute in the limit $\gamma \rightarrow \infty$, it doesn't do so for finite $ \gamma$, unlike in the 3+1 case. That $ \gamma$ does not induce canonical transformation in the ADM phase space also is clear in this context.

\subsection{The Singularity and its Resolution at $ \gamma \to  1$} \label{singular}
As it is apparent from the canonical structure, eg \eref{PB1}, \eref{PB2} the canonical structure blows up at the point $ \gamma \to \pm 1$. This is due to the fact that the lagrangian \eref{lag} and the action functional \eref{new} becomes independent of either of the 1-form fields $A^{\SR}$ for $ \gamma \to \pm 1$. As a result the symplectic structure we have constructed \eref{symp1}, becomes degenerate on the space $\mathcal V_{F}^{\SP}\times \mathcal V_{F}^{\SM}$ (leaving the gauge degeneracies apart), resulting it to be non-invertible. This is clearly the reason for blowing up of the equal time Poisson brackets \eref{PB1}.

In order to avoid this singularity we restrict our theory to $ \gamma \in \{ \mathbbm{R}^{+} - \{1\}\}$ and propose the theory \eref{new} for gravity in 2+1 dimensions. We will see further restriction on the range of $ \gamma$ is put by the quantum theory. The borderline case $ \gamma =1$ can however be dealt as follows. At the point $\gamma =1$ the effective theory of 2+1 gravity, as recovered from \eref{new} easily, is described by the single gauge 1-form $ A^{\SP}_I$ and we consider the phase space to be only coordinatized by flat connections $ A^{\SP}_I$ , ie $\mathcal V_{F}^{\SP}$ with the action functional:
\bea \label{effective}
I= \frac{l}{8 \pi G} \int _{ M}\left( A^{\SP I} \wedge d A^{\SP}_I + \frac{1}{3} \epsilon _{IJK} A^{\SP I} \wedge A^{\SP J} \wedge A^{\SP K} \right)
\eea
On the space $\mathcal V_{F}^{\SP}$ we now have the symplectic structure
\bea \label{sympf}
\Omega \left( \delta _1, \delta _2\right) =\frac{l}{4\pi G} \int _{\Sigma} \delta _{[1} A^{\SP I} \wedge \delta _{2]} A^{\SP}_I
\eea
This gives the non-singular Poisson bracket:
\bea \label{PBf}
\{A^{\SP I}_i (x,t),A^{\SP J}_j (y,t)\}= \frac{4\pi G }{l} \varepsilon _{ij} \eta ^{IJ} \delta ^2 \left(x,y \right)
\eea
In a more generalized theory, such as cosmological topologically massive gravity dealt in the first order formalism \cite{Grumiller:2008pr},\cite{Blagojevic:2008bn},\cite{Blagojevic:2009ek} one deals with the action
\bea \label{ctmg1}
I_{\scriptsize{\mbox{CTMG}}} &=& \frac{l}{16\pi G}\left[ \left(1/\gamma +1 \right) I^{\SP} + \left(1/\gamma -1 \right) I^{\SM} + \varrho ^{I} \wedge \left( d e_I + \epsilon _{IJK} e^J \wedge \omega ^K \right)\right]\nonumber \\
&=& \frac{l}{16\pi G}\bigg[ \left(1/\gamma +1 \right) I^{\SP} +  \left(1/\gamma -1 \right) I^{\SM} +\frac{1}{4l}\int_{M}\varrho ^{I} \wedge \left( d A^{\SP}_I + \epsilon _{IJK} A^{\SP J} \wedge A^{\SP K} \right)\nonumber\\&&- \frac{1}{4l}\int_{M}\varrho ^{I} \wedge \left( d A^{\SM}_I + \epsilon _{IJK} A^{\SM J} \wedge A^{\SM K} \right)\bigg] 
\eea
where $ \varrho$ is a new 1-form field which enhances the covariant phase space and emerges as a lagrange multiplier. The correponding symplectic structure is 
\bea \label{ctsymp}
\Omega _{ \scriptsize{\mbox{CTMG}}} \left( \delta _1, \delta _2\right)&=& \frac{l}{8\pi G}\bigg[\left(1/\gamma + 1 \right)\int _{\Sigma} \delta _{1} A^{\SP I} \wedge \delta _{2} A^{\SP}_I \nonumber\\&+& \left(1/\gamma - 1 \right)\int _{\Sigma} \delta _{1} A^{\SM I} \wedge \delta _{2} A^{\SM}_I  - \frac{1}{2l} \int _{\Sigma} \left(\delta _{[1} \rho ^I \wedge \delta _{2]} A^{\SP}\right)\nonumber \\&+& \frac{1}{2l} \int _{\Sigma} \left(\delta _{[1} \rho ^I \wedge \delta _{2]} A^{\SM}\right)\bigg]
\eea
In contrast to the theory we have considered, this theory does not become independent of any of the dynamical variables $ \left( A ^{\SM}, A ^{\SP} , \varrho\right)$ as $ \gamma \rightarrow 1$:
\bea \label{ctmg2}
I_{\scriptsize{\mbox{CTMG}}} \bigg{|}_{ \gamma =1} &=& \frac{l}{16\pi G}\bigg[ 2 I^{\SP} + \frac{1}{4l}\int_{M}\varrho ^{I} \wedge \left( d A^{\SP}_I + \epsilon _{IJK} A^{\SP J} \wedge A^{\SP K} \right)\nonumber\\&&- \frac{1}{4l}\int_{M}\varrho ^{I} \wedge \left( d A^{\SM}_I + \epsilon _{IJK} A^{\SM J} \wedge A^{\SM K} \right)\bigg] 
\eea and in this limit $ \gamma \to 1$ the symplectic structure \eref{ctsymp} remains non-degenerate:
\bea
\Omega _{ \scriptsize{\mbox{CTMG}}}\bigg{|}_{ \gamma =1} \left( \delta _1, \delta _2\right)&=& \frac{l}{8\pi G}\bigg[2\int _{\Sigma} \delta _{1} A^{\SP I} \wedge \delta _{2} A^{\SP}_I  - \frac{1}{2l} \int _{\Sigma} \left(\delta _{[1} \rho ^I \wedge \delta _{2]} A^{\SP}\right)\nonumber \\&+& \frac{1}{2l} \int _{\Sigma} \left(\delta _{[1} \rho ^I \wedge \delta _{2]} A^{\SM}\right)\bigg]
\eea
On the other hand there is a price one has to pay for considering TMGs in general. The theory develops a local propagating degree of freedom (graviton) and the complete non-perturbative quantization (as we present in the next section) seems far from being a plausible aim. Progress in perturbative quantization about linearized modes in TMG although have been made and relevance of the limit $ \gamma \rightarrow 1$ in this context made clear in \cite{Li:2008dq}.
\subsection{The physical phase space}\label{phys}
The route we choose for quantization of the system involves eliminating the gauge redundancy inherent in the theory, ie, finding the solution space modulo gauge transformations, in the classical level itself. For the present purpose this approach is useful in contrast to the other one which involves quantizing all degrees of freedom and then singling out the physical state space as the solution of the equation:
$$\mathbf{\hat M| \Psi \rangle }= 0$$
ie, the kernel of the quantum version of the constraints or the master constraint (regularized suitably). For illustrations of this later path one may look up the context of quantization of diffeomorphism invariant theories of connections \cite{Ashtekar:1995zh}, eg, loop quantum gravity in 3+1 dimensions \cite{Ashtekar:2004eh}.

The advantage of the first approach, ie, the reduced phase space (constrained first) one is that the phase space is completely coordinatized by gauge invariant objects; another manifestation being its finite dimensionality. Quantization of a finite dimensional phase space may acquire non-triviality only through the topology of it, as will be illustrated in the case at hand. 

Now, the physical phase space is clearly $$\left( \mathcal V_{F}^{\SP}/ \sim \right)\times \left(\mathcal V_{F}^{\SM}/ \sim \right),$$ where $\sim$ means equivalence of two flat connections which are gauge related. It is thus understood \cite{Witten:1988hc} that at least for the case when $ \Sigma$ is compact, each of the $\mathcal V_F ^{\SR} / \sim $ spaces is topologically isomorphic to the space  $\left( \hom : \pi _1 (\Sigma) \rightarrow SO(2,1) \right)/\sim)$ of homomorphisms from the first homotopy group of $\Sigma$ to the gauge group modulo gauge transformations. This isomorphism is realized (parameterized) by the holonomies of the flat connections around non-contractible loops on $ \Sigma$ which serve as the homomorphism maps.

For the choice of the topology of compact $\Sigma$, one may start by choosing a general $g$-genus Riemann surface. The case $g=0$ is trivial, and the moduli space consists of two points. For $g \geq 2$, parametrization of the phase space is highly non-trivial and topology of it is still not clear in literature, although construction of canonical structure on those moduli spaces have been constructed \cite{Nelson:1989zd}. As the first non-trivial case we therefore choose the case when $ \Sigma$ is a genus 1 Riemann surface $T^2$. For this torus, we know that $ \pi_1 (T^2) = \mathbbm{Z} \oplus \mathbbm{Z}$ i.e. this group is freely generated by two abelian generators $ \alpha$ and $ \beta$ with the relation 
\bea \label{relation} 
 \alpha \circ \beta =\beta \circ \alpha .
\eea
Since the connections at hand are flat, their holonomies depend only upon the homotopy class of the curve over which the holonomy is defined. For this reason, as parameterizations of the $ \mathcal V_{F}^{\SR}$ we choose the holonomies %\cite{Ezawa:1994nv}
$$h^{\SR}[ \alpha]:= \mathcal{P} \exp \left(\int _{ \alpha} A ^{\SR}\right)~~\mbox{ and}~~h^{\SR}[ \beta]:= 
\mathcal{P} \exp \left(\int _{ \beta} A ^{\SR}\right)$$
\footnote{here the path ordering $ \mathcal{P}$ means ordering fields with smaller parameter of the path to the left}with \eref{relation} being implemented on these $SO(2,1)$ group valued holonomies as 
\bea \label{relation2}
h^{\SR}[ \alpha] h^{\SR}[ \beta] &=& h^{\SR}[ \beta]h^{\SR}[ \alpha].
\eea
As is well-known these are gauge covariant objects although their traces, the Wilson loops are gauge invariant. 
%Suppose one has $c^{\SR} [ \alpha ] $ and $c^{\SR}[ \beta]$ as the traces of  $h^{\SR}[ \alpha]$ and $h^{\SR}[ \beta]$ above. Then following \cite{Nelson:1989zd} one has the following Poisson bracket induced by \eref{PB1}\bea \label{PB4}\{c^{\SR}[\alpha], c^{\SR}[ \beta]\} = gr\eea
Although the classical Poisson bracket algebra of Wilson loops for arbitrary genus were exhaustively studied in \cite{Nelson:1989zd}, the phase space these loops constitute is absent. On the other hand there is another simple way of finding the gauge invariant space especially for the case of genus 1, as outlined in \cite{Ezawa:1994nv}, \cite{Imbimbo:1991he}. We will for completeness briefly give the arguments reaching the construction.

Under the gauge transformations $A^{\SR} \rightarrow \tilde{A^{\SR}} = g^{-1}\left( A^{\SR}+d\right)g$ the holonomies transform as $h^{\SR}[c] \rightarrow \tilde{h}^{\SR}[c] = \chi ^{-1} h^{\SR}[c] \chi$ for any closed curve c and some element $\chi \in SL(2, \mathbbm{R})$.

Again from \eref{iwasawa} we know that any $SO(2,1)$ element is conjugate to elements in any of the abelian subgroups: $f_{\phi}$ or $g_{\xi}$ or $h_{\eta}$. Out of the three cases, for illustrative purpose we present the elliptic case.

Let $h^{\SR}[\alpha]$ is conjugate to an element in the elliptic class. Up to proper conjugation we can write $$h^{\SR}[\alpha]= e^{-\lambda _0 \rho _{\SR}}$$ and from the discussion of \ref{app} with \eref{relation2} we must have that
$$h^{\SR}[\beta]= e^{-\lambda _0 \sigma _{\SR}}.$$
Hence we have $\rho_{\SR}$ and $ \sigma _{\SR}$ with range $(0,2\pi)$ parameterizing a sector of the gauge invariant phase space with topology of a torus : $S^1 \times S^1 \simeq T^2$. 

Similarly structures of the other two sectors can also be found out. One is $\left(\mathbbm{R}^2 \backslash \{0,0\}\right)/\mathbbm{Z}^2$, containing an orbifold singularity and another is $S^1$ topologically. The total phase space is therefore product of two identical copies of $T^2 \cup \left(\mathbbm{R}^2 \backslash \{0,0\}\right)/\mathbbm{Z}^2 \cup S^1$. To be more precise the total phase space can be thought of as a union of a punctured torus $\tilde T$, the punctured orbifold $\left(\mathbbm{R}^2 \backslash \{0,0\}\right)/\mathbbm{Z}^2$ named as $\tilde P$ glued together at the repsective punctures through the circle $S^1$, identifying the $S^1$ as a point. 

%Thus the total phase space has in total $3^2=9$ sectors. But there is a-priori no reason for them not to be disjoint sets. There also is no obvious reason for making an artificial gluing of the three sectors of either $+$ or $-$ kind as done in \cite{Ezawa:1994nv}. For the purpose of the present paper we will confine us to the sector constructed above and quantize this part of the phase space only. Possibility of carrying out a quantization programm on these sectors is discussed later. At the quantum level there of course may be non-trivial tunnelling amplitudes between these disjoint sectors.

\subsection{Symplectic structure on the phase space}
If one considers periodic coordinates $x, y$ on $\Sigma \simeq T^2$ with period 1, then it follows immediately that the connections 
\bea \label{conn}
A^{\SR} = \lambda _0 (\rho _ {\SR} dx + \sigma _{\SR} dy)
\eea
 give the above written holonomies parameterizing the $ \tilde{ T}$ sector.

%\begin{figure}
%\centering
%\includegraphics[height=55mm,width=70mm]{chhobi.png}
%\caption{Projection from the pre-symplectic manifold $ \left(\mathcal{V}_F^{\SP}\times\mathcal{V}_F^{\SM}, \Omega\right)$ to the gauge invariant one.}
%\end{figure}

Now using \eref{symp1} and \eref{symp2} we have the symplectic structure $\omega$, whose pull back to the pre-symplectic manifold is $ \Omega$ on this $ \tilde{T}$ sector of the phase space is given by:
\bea \label{symp3}
\omega \left( \delta _1 , \delta _2\right) = \frac{l}{4\pi G} \left[ \left(1/\gamma + 1\right) \delta _{[1}\rho _{\SP} \delta _{2]} \sigma _{\SP}+ 
\left(1/\gamma - 1\right) \delta _{[1}\rho _{\SM} \delta _{2]} \sigma _{\SM}\right]
\eea 
or,
\bea \label{symp4}
\omega &=& \frac{l}{4\pi G} \left[ \left(1/\gamma + 1\right){ \bf{d}}\rho _{\SP}\wedge{\bf{d}} \sigma _{\SP}+\left(1/\gamma - 1\right){ \bf{d}}\rho 
_{\SM} {\wedge}{ \bf{d}}\sigma _{\SM}\right] \nonumber\\
&=& \frac{k_{\SP}}{2\pi} {\bf{d}}\rho_{\SP} \wedge {\bf{d}}\sigma_{\SP}+\frac{k_{\SM}}{2\pi} {\bf{d}}\rho_{\SM} \wedge {\bf{d}}\sigma_{\SM}
\eea
where $k_{\SR} = \frac{l \left(1/\gamma \pm 1 \right)}{2G}$ and the $\bf d$ are exterior differentials on the phase and the $\wedge$ is also on this manifold, not on space time. 
Here  we  introduce  holomorphic  coordinates on $ \tilde{T}$ corresponding to a complex structure $\tau$ on the two dimensional space manifold $\Sigma$ as 
$${z_{\SR}} = \frac{1}{\pi}\left({\rho _ {\SR} } + \tau \sigma _{\SR} \right).$$ 
Then the symplectic structure in \eref{symp4} takes the form
\bea \label{symp5}
\omega = \frac{ik_{\SP} \pi}{4\tau _2} {\bf{d}}z_{\SP} \wedge {\bf{d}}\bar z_{\SP}+\frac{ik_{\SM} \pi}{4\tau _2} {\bf{d}}z_{\SM} \wedge {\bf{d}}\bar z_{\SM}
\eea
In a similar fashion the symplectic structure on $ \tilde P$ is given by:
\bea \label{symp6}
\omega = \frac{ik_{\SP} \pi}{4\tau _2} {\bf{d}}z_{\SP} \wedge {\bf{d}}\bar z_{\SP}+\frac{ik_{\SM} \pi}{4\tau _2} {\bf{d}}z_{\SM} \wedge {\bf{d}}\bar z_{\SM}
\eea
where $z_{\SR} = \frac{1}{\pi}\left(x _{\SR}+\tau y_{\SR}\right)$, $x, y$ being the coordinates on $\tilde P$.
\section{Geometric quantization of the phase space}\label{geo_q}
As explained in \ref{phys} the total phase space is product of two identical copies of $\tilde T \cup \tilde P$, $\tilde T$ and $ \tilde{P} $ being glued through a circle $S^1$ around the puncture at $(0,0)$. Variables relevant to each factor of this product has been distinguished until now by $\pm$ suffices. From now on, we will remove this distinction for notational convenience and will restore when it is necessary. 

Upon quantization the total wave functions (holomorphic sections of the line bundle over $ \tilde{T} \cup \tilde{P}$) should be such that the wave function (holomorphic sections of the line bundle over $ \tilde{T} \cup \tilde{P}$) on $\tilde T$, say $\psi (z)$ and the wave function on $\tilde P$, say $\chi (z)$ should `match' on the circle. The plan of quantization is therefore simple. We will first carry out the quantization on $ \tilde{T} $. Then we will consider those functions on $ \tilde{P}$ which can be found by continuation in some sense of the wave functions on $ \tilde{T} $ .
\subsection{Quantization on $\tilde T$} \label{geqT}
While performing quantization on $ \tilde{T}$ with the symplectic structure 
$$ \omega = \frac{k}{2 \pi} {\bf d} \rho \wedge {\bf d} \sigma =\frac{ik \pi}{4 \tau _2}{\bf d} z \wedge {\bf d} \bar z$$
one must keep in mind the fact that $ \tilde{T} $ is in fact punctured as opposed to being compact. \footnote{Had the symplectic manifold $ \tilde{T} \cup \tilde{P} $ been compact, Weil's integrality criterion would require the Chern-Simons level $k$ to be integer valued. At this point we keep open the possibility of $k$ being any real number.} The distinction occurs from the non triviality of the algebra of the generators of the homotopy group. The three generators of $ \pi _1 ( \tilde{T}) $, denoted as $a,b, \& \Delta$ respectively correspond to the usual cycles of the compact torus and the cycle winding around the puncture. They should satisfy the following relations:
$$aba^{-1}b^{-1} = \Delta \qquad a\Delta a^{-1} \Delta ^{-1} =1 \qquad b\Delta b^{-1} \Delta ^{-1} =1$$
As explained in  \cite{Imbimbo:1991he},\cite{'tHooft:1977hy} $q \in \mathbbm{Z}$ dimensional unitary representation of these relations are given as follows. The unitary finite dimensional non-trivial representations of this algebra must have the commuting generator $ \delta $ proportional to identity. Hence we have the for some $q$ dimensional representation
$$ \Delta _{\alpha ,\beta} = \e ^{2\pi i p/q} \delta _{\alpha ,\beta}$$ where $p, q$ are positive integers, co prime to each other. Reason behind choosing rational phase will become clear shortly when we complete the quantization. 

Again, up to arbitrary $U(1)$ phase factor $a, b$ are represented as
$$ a _{\alpha ,\beta} = \e ^{-2 \pi i \frac{p}{q} \alpha} \delta _{\alpha ,\beta} \qquad b _{\alpha ,\beta} = \delta _{\alpha ,\beta +1}$$ with $\alpha, \beta \in \mathbbm{Z}_q$. It is also being expected that the space of holomorphic sections should also carry the $q$ representation of this homotopy group.

Let us now consider quantization on $ \mathbbm{R}^2$ endowed with complex structure $\tau$ and the above symplectic structure. The fact that the actual phase space we wish to quantize is a punctured torus will be taken into account by action of the discretized Heisenberg group operators on the Hilbert space of parallel sections of the line bundle over $ \mathbbm{R}^2$. A very similar quantization scheme for a different situation may be found in \cite{Thiemann:2007zz}, \cite{Ashtekar:2000eq},\cite{Kloster:2007cb}. 

Start from the symplectic structure on $ \mathbbm{R}^2$ instead of the the punctured torus $\tilde T$ and coordinatize it by $\rho$ and $\sigma$ and endowed with complex structure $ \tau$, such that holomorphic anti holomorphic coordinates are chosen as before: 
$$\omega = \frac{k}{2\pi} {\bf d} \rho \wedge {\bf d} \sigma .$$ With definition of holomorphic coordinate $z = \frac{1}{\pi}( \rho + \tau \sigma)$ defined through arbitrary complex structure $\tau$ this becomes 
$$ \omega = \frac{ik\pi}{4 \tau _2} {\bf d}z \wedge {\bf d} \bar{z}, $$ where $\tau _2 = \Im \tau.$
It is easy to check that the symplectic potential
$$ \Theta = \frac{ik\pi}{8\tau _2}\left[- \left( \bar{z} - 2z \right){\bf d}z + \left( z + \xi ( \bar z)\right) {\bf d} \bar{z}\right]$$ gives the above symplectic structure, for arbitrary anti-holomorphic function $\xi (\bar{z})$.
Let us now consider the hamiltonian vector fields corresponding to the variables $ \rho$ and $ \sigma$
\bea \label{hamil_vec_f}
\zeta _{\rho} = \frac{2\pi}{k} \partial _{\sigma} \\
\zeta _{\sigma} = -\frac{2\pi}{k} \partial _{\rho}
\eea
The corresponding pre-quantum operators to these variables are therefore
\bea \label{rho}
\hat{ \rho} &=& - i \zeta _{\rho} - \Theta (\zeta _{\rho}) + \rho \nonumber\\
            &=& - \frac{2i}{k} \left( \tau \partial _z + \bar \tau \partial _{\bar z} \right) + \frac{i\pi}{4 \tau _2} \left( \bar{\tau} z - \tau \bar{z} - 2 \tau z - \bar \tau \xi (\bar z)\right) \\
\hat{ \sigma} &=& - i \zeta _{\sigma} - \Theta (\zeta _{\sigma}) + \sigma \nonumber\\
            &=& - \frac{2i}{k} \left( \partial _z + \partial _{\bar z} \right) + \frac{i\pi}{4 \tau _2} \left(  z + \bar{z} + \xi (\bar z)\right)
\eea
Now, parallel (holomorphic) sections of the line bundle $ \pi : L \tilde{T} \rightarrow \tilde{T}$ over the symplectic manifold $\tilde T$ are classified through the kernel of the Cauchy-Riemann operator defined via the connection $ \nabla = {\bf d}-i \Theta$ on $L \tilde{T}$ as (in units of $\hbar$=1)
\bea \label{C-R}
\nabla _{ \partial _{ \bar z}} \Psi (z, \bar z) = 0 .
\eea
Ans\"atz for $\Psi$ can be chosen as:
\bea \label{ansatz}\Psi (z, \bar z) = \e^{-\frac{k \pi}{8 \tau _2} \left( z \bar{z} + \Xi( \bar z)\right) } \psi (z) \eea
with $ \Xi( \bar{z} ) $ being the primitive of $\xi (\bar{z} )$ with respect to $ \bar{z}$ and $\psi (z)$ is any holomorphic function. This is how the holomorphic factor $ \psi(z)$ of the function $ \Psi (z)$ is being singled out by the $ \nabla _{ \partial _{ \bar{z}}}$.
To find the representations of the operators corresponding to $\hat \sigma$ and $\hat \rho $ on the space of the holomorphic functions, we see the actions:
\bea
\hat{ \rho} \Psi (z, \bar z) &=&  \e^{-\frac{k \pi}{8 \tau _2} \left( z \bar{z} + \Xi( \bar z)\right) } \left[ - \frac{2i}{k} \tau \partial _z +\pi z \right] \psi(z) \nonumber\\
                             &=:& \e^{-\frac{k \pi}{8 \tau _2} \left( z \bar{z} + \Xi( \bar z)\right) } \hat{ \rho ^{\prime}} \psi(z) \\
\hat{ \sigma} \Psi (z, \bar z) &=&  \e^{-\frac{k \pi}{8 \tau _2} \left( z \bar{z} + \Xi( \bar z)\right) } \left[ \frac{2i}{k} \partial _z  \right] \psi(z) \nonumber\\
                             &=:& \e^{-\frac{k \pi}{8 \tau _2} \left( z \bar{z} + \Xi( \bar z)\right) } \hat{ \sigma ^{\prime}} \psi(z) .
\eea
 These give the representations for $ \sigma$ and $ \rho$ on the space of holomorphic sections in terms of $\hat{ \rho ^{\prime}}$ and $\hat{ \sigma ^{\prime}}$.
 
 At this point it is necessary to notice that we aim to quantize the punctured torus instead of $ \mathbbm{R}^2$. This is done by imposing periodicity conditions (for being defined on torus) through action of the Heisenberg group
and the homotopy group (accounting for the puncture) on the space of holomorphic sections. Let us therefore define homotopy matrix-valued Heisenberg operators:
\bea
 U(m) := b^{m} \e ^{ikm \hat{\rho ^{\prime}}}\\
 V(m) := a^{m} \e ^{- ikm \hat{\sigma ^{\prime}}}
\eea
The periodicity condition that,
$$U(m)V(n) \psi(z) = \psi (z)$$ for $m,n \in \mathbbm{Z}$ therefore reduces to
\bea \label{theta1}
\psi (z + 2m + 2n \tau)= \e ^{ - ikn^2 \pi \tau - ikn \pi z} a^{-m} b^{-n} \psi (z) \qquad \mbox{ or}\nonumber \\ 
\psi _{ \alpha}(z + 2m + 2n \tau)= \e ^{ - ikn^2 \pi \tau - ikn \pi z + 2\pi i (p/q)m \alpha} \psi _{ \alpha +n}(z).
\eea
 in terms of components.
%From this point onwards, we will concern ourselves with cases where $k$ is a positive rational number. By observation, components of the $q$ component functions obeying \eref{theta1} are given by nothing but level $k$, $SU$ theta functions \cite{mum}, depicted as:

Let us now as a digression concentrate upon level $I,J$  $SU(2)$ theta functions
$$ \vartheta _{I,J}( \tau, z) := \sum_{j \in \mathbbm{Z}} \exp \left[2\pi i J \tau \left( j + \frac{I}{2J}\right)^2 + 2\pi i J z \left(  j + \frac{I}{2J}\right)\right]$$% with the levels defined by $I= qN+ p\alpha$, $J=pq/2$ and $z$ replaced by $z/q$ transform as
and define
$$ \tilde{ \vartheta} _{ \alpha, N} (\tau,z):= \vartheta _{qN+ p\alpha,pq/2}( \tau, z/q) $$
for $pq$ even \cite{Imbimbo:1991he}.
After some manipulations, it is easy to check that 
\bea \label{theta2}
\tilde{\vartheta} _{\alpha, N}( \tau, z + 2m +2n \tau) = \e ^{-\pi i (p/q) n^2 \tau - \pi i (p/q) nz} \e ^{2\pi i (p/q) m \alpha} \tilde{\vartheta} _{ \alpha + n, N} ( \tau ,z)
\eea the indices $ \alpha \in \{ 0,1, \ldots ,q-1\}$ and $N \in \{0,1, \ldots ,p-1\}$. These theta functions are known to form a complete $p$ dimensional set over the field of complex numbers \cite{mum}. 

Again comparing the transformations \eref{theta1} and \eref{theta2} we infer that for the value $k =p/q$, \footnote{From another point of view it can be seen that the monodromy of wave functions about the puncture satisfying the above relation is measured to be $\e ^{2\pi i k}$. When this is related related with the measure of non-commutativity $\e ^{2 \pi i p/q}$ of the homotopy generators due to the puncture \cite{Imbimbo:1991he} we have the relation: $$k =p/q$$ up to additive integers. } a positive rational, we have a finite $p$ dimensional vector space of physical states spanned by $q$ component wave-functions, represented by theta functions depicted as above. For instance the $N$ th wave function is
$$ \psi ^{N} (z)= \left(\begin{array}{c} \tilde{ \vartheta} _{0,N} (\tau , z)\\ \vdots \\ \vdots \\\tilde{ \vartheta} _{q-1,N}( \tau ,z) \end{array}\right) .$$
Here we have only considered the case $pq$ even. In spirit the case $pq$ odd \cite{Imbimbo:1991he}can also be dealt at par. Distinction of that case from the present one occurs as identification of the wave functions satisfying \eref{theta1} has to be made with a theta functions with different levels. 
%$\psi (z)$ on $\tilde T$, according to the holomorphic quantization scheme (reference) should be $\psi _{\SP}(z _{\SP}) \otimes \psi _{\SM}(z _{SM})$ corresponding to the two parts of the symplectic structure $\Omega _{\SR}$ in \eref{symp5}. For brevity we keep the notation $\psi (z)$ for either of them corresponding to the syplectic structure 
%\bea \label{symp7} 
%\tilde{ \vartheta} _{1,N} (\tau ,z)\\ \vdots \\ \vdots \\ \tilde{ \vartheta} _{q,N}  (\tau, z)
%\Omega = \frac{ik\pi}{4\tau_2} {\bf d} z \wedge {\bf d} \bar z
%\eea
%On $\tilde T$ we know that the generators of the first homotopy group obey  the relations $aba^{-1}b^{-1} = \delta$, $a\delta a^{-1} \delta ^{1} =1$ and $b\delta b^{-1} \delta ^{1} =1$, where $\delta$ corresponds to a cycle surrounding the puncture $(0,0)$; $a, b$ being the usual two non-trivial cycles. In order to have a faithful non-trivial representations of the homotopy group, $a, b, \delta$ should be represented by matrices. Hence the wave function $\psi (z)$ shoud transform in some irreducible unitary q-dimensional representation of the non-abelian homotopy group $\pi _1 (T^2 \backslash \{0,0\}) \equiv \pi _1 (\tilde T)$.

%The wave function $\psi (z)$ turns out to have $q$-components satisfying (see appendix)
%$$\psi (z+2m+2n \tau) = \e^{- \pi i k \tau n^2 - \pi i k z n} a^m b^n \psi (z)$$
%where $a,b$ are $q\times q$ unitary matrices.
We are considering $k = p/q$, a positive rational. From our earlier discussions \eref{symp4}, we had $ k_{ \SR} = \frac{l ( 1/\gamma \pm 1)}{2G}$ (in the units of $\hbar =1=c$, and in 2+1 space time dimensions $G$ is of dimension that of length, hence making $k$ dimensionless) in terms of the parameters of the classical theory. From the point of view of quantization, we are restricting only those values of classical parameters for which the combinations $ k _{\SR}$ are positive rational.
\subsection{Continuation to $\tilde P$} \label{geoP}
%For the wave function $\chi (z)$ on $\tilde P$  we must be of the form
%$$\chi(z) = z^{\kappa} \phi (z)$$
%where $\phi (z)$ is holomorphic and $\kappa \in \mathbbm Q$. The factor $z^\kappa$ in the wave function is necessary since wave function must be allowed to pick up a non-trivial phase in going around the orbifold singularity ${0,0}$. Again since $\tilde P \equiv (\mathbbm {R}^2 \backslash {0,0})/\mathbbm {Z}_2$, the wave functions defined on it must have definite `parity'. As a result $\phi(z)$ must be even or odd.
The wave function $\chi (z)$ on $\tilde P$  must be of the form
$$\chi(z) = z^{\kappa} \phi (z)$$
where $\phi (z)$ is holomorphic and $\kappa $ is a positive rational. The factor $z^\kappa$ in the wave function is necessary since it must be allowed to pick up a non-trivial phase in going around the orbifold singularity. 

Also the wave function on the entire phase space should be such that the two functions $ \psi$ and $ \chi$ agree on the intersection and the wave function $\psi (z)$ on $\tilde T$ should uniquely determine that on $\tilde P$ in a neighbourhood of the intersection. Hence $\chi(z)$ must take the following form around the origin.
\bea \label{chi}
{\chi_N^{\alpha}}(z) = {\e^{\frac{2\pi i\alpha r}{q}}}{z^{\frac{r}{q}}} {\phi_N^{\alpha}} (z)
\eea
%Again since $\tilde P \equiv (\mathbbm {R}^2 \backslash {0,0})/\mathbbm {Z}_2$, the wave functions defined on it must have definite `parity'. As a result $\phi(z)$ must be even or odd.
In the above equation we have chosen $ \kappa = \frac{r}{q}$ keeping in mind that $ \chi ^{ \alpha}_N (z)$ should have exactly $q$ number of branches. This is necessary for agreement of $ \psi$ and $ \chi$ around the puncture. $z^{\frac{r}{q}}$ in \eref{chi} is the principal branch of $z ^{\kappa}$.

 Again since $\tilde P \equiv (\mathbbm {R}^2 \backslash {0,0})/\mathbbm {Z}_2$, the wave functions defined on it must have definite `parity' since this results into a constant phase factor in the wave function. As a result ${\phi_N^{\alpha}} (z) $ must be even or odd. This property must hold for the wave functions on $\tilde T$ in order that the wave functions agree on a circle around the origin. For example in the case $pq$ even \cite{Imbimbo:1991he} we construct from \eref{theta2} wave functions with definite parity through the combination:
\bea \label{parity_psi}
\psi_N^{\alpha \SR}(\tau ,z) = \tilde{\vartheta} _{\alpha, N} (\tau ,z)  \pm  \tilde{\vartheta} _{-\alpha, -N} (\tau ,z)  %\pm {\theta_{{-({qN+p\alpha})},{\frac{pq}{2}}}}(\tau ,{\frac{z}{q}}) 
\eea

Now in order to match the wave-functions, we have to do Laurent expansion around the origin. Laurent expansion about a point say $P_+$ on a torus was studied in \cite{Krichever:1987qk}. It was shown that there exists a basis  on $C_ \eta$ which is anlogous to the basis $z^{n}$ on a circle, where $C_\eta$ parametrizes a compact Riemann surface in same way as a circle can parametrizes the extended complex plane, $ \eta $ being a well defined global parameter that labels the curve $C_\eta = \{Q : \Re [ p(Q)] \}= \eta$, $p(Q) = {\int_{Q_0}^{Q}} dp$ , $dp$ on the other hand is a differential of third kind on the Riemann surface with poles of the first order at the points $P_{\pm}$ with residues $\pm 1$. In the case of a torus an exact basis $A_n (z)$ on  $C_\eta$ is given in \cite{Krichever:1987qk}. These are the Laurent basis for curves on the torus on a special system of contours $C_\eta$ . As $\eta \rightarrow \pm \infty$, $C_\eta$ are small circles enveloping the point , ${P_{\mp}}$ . We have to match the wave function on the the torus around a small circle about $P_+$ with that on $\tilde P$. We could have expanded of $\psi (z) $ in terms of the basis $A_n (z)$  while for the latter we can expand $\chi (z)$ in terms of $z^n$. The two expressions must be equal, when an expansion of the basis $A_n (z)$ is performed in terms of $z^n$. Since $\psi_N^{\alpha \SR}(\tau ,z)$ is holomorphic we have the Laurent expansion for $ \phi_N^{\alpha \SR} ( \tau,z)$, (which is related to $ \chi _N ^{\alpha \SR} ( \tau,z)$ through \eref{chi}) around the origin as follows:
\bea \label{chi2}
\hspace{-2.0cm}\phi _N^{\alpha \SR}(\tau ,z) = 1 + \left[ \frac{u^2}{2!} \left( \pi i pq\right) ^{-1} \partial _{ \tau} +  \frac{u^4}{4!} \left( \pi i pq\right) ^{-2} \partial _{ \tau}^2 + \cdots \right]  \times \sum _j  \left( \e^{ \pi i pq \tau x_j ^2} \pm \e^{ \pi i pq \tau {\tilde x_j}^2}\right)  \nonumber\\
\hspace{-0.62cm}+  \left[ u+ \frac{u^3}{3!} \left( \pi i pq\right) ^{-1} \partial _{ \tau} +  \frac{u^5}{5!} \left( \pi i pq\right) ^{-2} \partial _{ \tau}^2 +\cdots \right]  \times \sum _j  \left(x_j \e^{ \pi i pq \tau x_j ^2} \pm {\tilde x_j}\e^{ \pi i pq \tau {\tilde x_j}^2}\right) 
\eea
with $u= i\pi pz$ and $ x_j = j+ \frac{qN+ p \alpha}{pq}$ and $ {\tilde x}_j = j- \frac{qN+ p \alpha}{pq}$.
$ \phi _N ^{ \alpha}$ in the wave function \eref{chi} should have the same form as above \eref{chi2}. This does not determine the exact form of the above function on the entire $ \tilde{P}$. But this asymptotic form on $ \tilde{P}$ ensures the finite number $p$ of the wave functions each with $q$ components.

Hence we have at hand the full Hilbert space of the quantized theory. Dimension of the Hilbert space is $ p_{\SP} p _{\SM}$. The extensions determined by the above asymptotic form should also be `square integrable' with respect to some well-defined measure $ {\bf d} \mu _{\tilde{P}}$.
 The unitarily invariant, polarization independent inner product associated with this Hilbert space of wave functions (to be more precise `half-densities') is given in temrs of the K\" ahler potential on $ \tilde{T}$ corresponding to \eref{symp5} or \eref{ansatz} and measure $ {\bf{d}} \mu _{ \tilde{P}}$  on $ \tilde{P}$ is given as:
$$ \langle \Psi , \Psi ^{ \prime}\rangle = \int _{ \tilde{T}} \sum _{ \alpha} {\bf d}z { \bf d} \bar{z} \tau ^{-1/2}_2 \e ^{ - \frac{k \pi}{8 \tau _2} \left( 2 z\bar{z}  + \Xi (z) + \Xi(\bar z)\right)} \psi ^{ \prime}_{\alpha}(z) \psi _{\alpha}(\bar{z}) + \int _{ \tilde{P}} \sum _{ \alpha} {\bf d} \mu _{ \tilde{P}} \chi ^{ \prime}_{\alpha} (z) \chi _{ \alpha} ( \bar{z})$$
\subsection{$ \gamma \rightarrow 1$ limit in quantum theory}
It dates back to Brown and Henneaux \cite{Brown:1986qj}, who first showed the existence of a pair of (identical) centrally extended Virasosro algebras as the canonical realizations of the asymptotic symmetris for 2+1 Einstein gravity with negative cosmological constant on an asymptotically AdS manifold. Later various authors \cite{Blagojevic:2005hd} for example, reproduced the result with equivalent theories of \eref{mielke} or toplogically massive gravity (TMG) \cite{Blagojevic:2008bn}, \cite{Kraus:2006kq}, \cite{Skenderis:2009nt} confirming an AdS(3)/CFT(2) correspondence, although with unequal central charges. In the theory we are dealing with, these central charges come out to be $ \left( c_{\SP}, c_{\SM}\right)= \frac{3l}{2G} \left( \left( 1+ \frac{1}{ \gamma}\right), \left( 1- \frac{1}{ \gamma}\right)\right)$ in our conventions and notations.

The chiral limit ie $ \gamma \to 1$ in this direction has gained importance in recent literature for various reasons. In view of results from \cite{Li:2008dq}, where second order TMG was studied on an asymptotically AdS spacetime, we see that in order to make sense of all the graviton modes $ \gamma$ should be restricted to 1. At this limit the theory becomes chiral with $\left( c_{\SP},c_{\SM}\right)= \left( \frac{3l}{G},0 \right)$. Another interesting result by Grumiller et al \cite{Grumiller:2010rm} reveals that at the quantum level chiral limit of TMG is good candidate as a dual to a logarithmic CFT (LCFT) with central charges $ \left( c_{\SP},c_{\SM}\right)= \left( \frac{3l}{G},0 \right)$. More recent works with some of the interesting ramifications of TMG `new massive gravity' \cite{Bergshoeff:2009hq} shows similar progress \cite{Grumiller:2009sn}. These results were worked out on an asymptotically AdS space-time. In the present case however, we have considered spatial slice to be a genus 1 compact Riemann surface, without boundary. Hence chance of a CFT living at the boundary doesn't arise. Even if we had worked on a asymptotically AdS manifold, the theory would not be dual to an LCFT, because for that a propagating degree of freedom is necessary, which is absent in our case.

However there are some interesting issues in the present discussion for the limit $ \gamma \to 1$:\\ We have inferred in \ref{geqT} from \eref{symp4},\eref{theta1} and \eref{theta2} $ k_{\SR} = \frac{l}{2G}(1/\gamma \pm 1)$, which are related to above discussed central charges through $k_{\SR} = \pm \frac{1}{3} c_{ \SR} $  must be positive rationals. As a result, if the ratio of the AdS radius $l$ and and Planck length $G$ (in units of $ \hbar=1=c$) is positive, we must restrict $ 0 < \gamma <1$. This is in apparent contradiction to the restriction $ \gamma \geq 1$ \cite{Solodukhin:2005ah} put by the CFT (living in the boundar, in the case of asymptotically AdS formulation). But this may well be resolved from the point of view that our analysis is completely performed on spacetime topology (as seen clearly in the construction of the physical phase space) whose spatial foliations are compact tori and relevant ranges of $ \gamma$ should depend non-trivially on the topology of spacetime and in our case restrictions coming from suitable CFT is not clear as explained in next paragraph.

As argued in \ref{singular} at the point $ \gamma =1$, we describe 2+1 gravity with negative cosmological constant through a single $SO(2,1)$ Chern Simons action \eref{effective}. On the other hand, for a rational $SO(2,1)$ (or any of its covers) Chern Simons theories on genus-1 spatial foliation, existence of a dual CFT  too is still not very clear, as argued in \cite{Imbimbo:1991he}. The modular transformation ($ SL(2, \mathbbm{Z})$) representations acting on the physical hilbert space (as found in \ref{geqT}, \ref{geoP}) appaear to be one of the two factors in to which modular representations of the conformal minimal models factorize. This observation points that a 2-D dual theory may not be conformal, although one may identify conformal blocks (of a CFT, if it exists) labelling our wavefunctions \cite{Imbimbo:1991he}.

\subsection{ Results on the quantization of parameters}
We have explained in section \ref{geqT}  that $ k _{ \SR}= \frac{ p _{\SR}}{q_{\SR}}$ are positive rationals. In \cite{Witten:2007kt} it has been shown that for the gauge group being an  n-fold diagonal cover of $SO(2,1) \times SO(2,1)$ ,  one requires the couplings 
\bea \label{qparam}
k_{\SP}\in 8 n^{-1}\mathbbm{Z} ~~\mbox{ for } n \mbox{ odd} \nonumber\\
k_{\SP} \in  4 n^{-1} \mathbbm{Z} ~~ \mbox{ for } n \mbox{ even  and} \nonumber\\
k_{\SP} + k_{\SM} \in 8 \mathbbm{Z}
\eea
 in our notation and convention. This is in agreement with our finding that the consistent quantization procedure reveals $k_{\SR} = \frac{p_{\SR}}{q_{\SR}} \in \mathbbm{Q}^{+}$ and we are considering $ q_{\SR}$ covers of the phase space (see section \ref{geoP}) which is constructed from the gauge group. In terms of physical parameters we have 
 \bea
\frac {l}{G}\in \mathbbm{Q}^+ ~~\mbox{ and}\nonumber\\
\frac {l}{G\gamma}\in \mathbbm{Q}^+ \nonumber\\
\Rightarrow \gamma \in \mathbbm{Q}^+
 \eea
 which are slightly less restrictive than the results of the analysis done in \cite{Witten:2007kt} $\frac {l}{G}\in \mathbbm{Q}^{+}$ and    $\frac {l}{G\gamma}\in \mathbbm{N} \subset \mathbbm{Q}^+$.

\section{Conclusion}
The features which come out of our analysis can be summarized as follows.

Classically it is observed that $ \gamma$ fails to induce canonical transformations on the canonical variables although equations of motion do not involve $ \gamma$. The role of $ \gamma$ is best viewed in the constraint strucure of the theory which is also studied in detail. On the other hand the `chiral' limit relevant in our case is $ \gamma \to 1+$ as opposed to the TMGs on asymptotically AdS space times, where it is $ \gamma \to 1-$. In the canonical structure the apparent singularity can also be removed as discussed in \ref{singular}.

Natuarally different values of $ \gamma$ results in inequivalent quantizations of the theory. Dimensionless $ \gamma$ and the cosmological constant $-\frac{1}{l^2}$ give the dimensionality of the physical state space in a subtle manner. Note that we had $k_{\SP} k_{\SM}=l^2 \frac{1/{\gamma ^2} - 1}{4 G^2}$, $k_{\SR} = \frac{p _{\SR}}{q _{\SR}}$, $p_{\SR}$ and $q_{\SR}$ being both positive integers and prime to each other. Dimension of the Hilbert space turns out to be $p _{\SP}p _{\SM}$ which must be a positive integer .This requirement, provides allowed  values of  $ \gamma$, for a given  $\frac {l}{2 G}$  such that   $\frac {l}{2G}\in \mathbbm{Q}^+$ and $\frac {l}{2G\gamma}\in \mathbbm{Q}^+$ .                   

\appendix 
\section{Conjugacy classes of $SL(2, \mathbbm{R})$}\label{app}
Any $SL(2, \mathbbm{R})$ (which is the double cover of $SO(2,1)$\footnote{Since from gravity action we got a 
gauge theory with a lie algebra shared commonly by $SO(2,1)$, $SL(2, \mathbbm{R})$, $SU(1,1) $ or any covering of 
them, the actual group used is quite irrelevant unless one is considering transformations between disconnected 
components of the group manifold.}) element $G$ can be written in its defining representation as the product of three 
matrices by the Iwasawa decomposition uniquely
\bea \label{iwasawa}
\hspace{-0.5cm}G = \underbrace{\left( \begin{array}{cc} \cos (\phi /2) & \sin (\phi /2) \\ -\sin (\phi /2) & \cos 
(\phi /2)\end{array}\right)}_{f _{\phi}} 
\underbrace{\left( \begin{array}{cc} e^{\xi /2} & 0 \\ 0 & e^{\xi /2}\end{array}\right)}_{g_\xi} 
\underbrace{\left( \begin{array}{cc} 1 & \eta \\ 0 & 1 \end{array}\right)}_{h_{\eta}}
\eea
with the range of $\phi$ being compact $(-2 \pi,2\pi)$ and those of $\xi$ and $\eta$ noncompact. Note that these 
three matrices fall in respectively the elliptic, hyperbolic and the null or parabolic conjugacy class of 
$SL(2, \mathbbm{R})$, in addition to forming three abelian subgroups themselves. Also note that
\bea
f_{\phi} &=& \exp \left( i \sigma _2 \phi /2\right) = e^{-\lambda _0 \phi}\nonumber \\
g_{\xi}  &=& \exp \left( \sigma _3 \xi /2\right) = e^ {\lambda _2 \xi}\nonumber \\
h_{\eta} &=& \exp \left[ (i \sigma _2 + \sigma _1) \eta /2\right] = e^{(-\lambda _0 -\lambda _1)\eta}  \nonumber
\eea
where $ \lambda _I \in \mathfrak{sl}(2,\mathbbm{R})$ with $\left[ \lambda _I, \lambda _J\right] = \epsilon _ {IJK} \lambda 
^K$.

We now state an important result which is used in the text. Let $g = \exp \left( \kappa _I \lambda ^I\right)$ and 
$g ^{ \prime} = \exp \left( \kappa ^{\prime}_{I} \lambda ^I\right)$ be two $SL(2, \mathbbm{R})$ elements. Then the 
necessary and sufficient condition for $g_1 g_2  = g_2 g_1$ to hold is $ \kappa^I = c \kappa^{ \prime I}$ for $I= 
0,1,2$ and any $c \in \mathbbm{R}$. This can be seen by using the Baker Campbell Hausdorff formula.

\section*{Acknowledgments}
The authors thank Parthasarathi Majumdar for suggesting investigations on this problem and for making numerous useful remarks and comments on the manuscript. One of the authors (RB) thank Council for Scientific and Industrial Research (CSIR), India for support through the SPM Fellowship (SPM-07/575(0061)/2009-EMR-I).

\end{document}